\documentclass{elsart}

\usepackage{graphicx,natbib,amssymb}

\journal{Advances in Space Research}
\begin{document}

\begin{frontmatter}



\title{2D stationary resistive MHD flows: borderline to magnetic reconnection solutions}


\author[Utrecht]{D.H. Nickeler\corauthref{cor}},
\corauth[cor]{Corresponding author.}
\ead{D.H.Nickeler@phys.uu.nl}
\author[Bonn]{H.-J. Fahr}
\ead{hfahr@astro.uni-bonn.de}

\address[Utrecht]{Astronomical Institute, Utrecht University, Princetonplein 5, 3584 CC Utrecht, 
the Netherlands}
\address[Bonn]{Institut f\"{u}r Astrophysik und Extraterrestrische Forschung, Auf dem H\"{u}gel 71, 
53121 Bonn, Germany}
 
\begin{abstract}
We present the basic equations for stationary, incompressible resistive
MHD flows in two dimensions. This leads to a system of differential equations
for two flux functions, one elliptic partial differential equation
(Grad-Shafranov--like) for the magnetic flux function and one for the
stream function of the flow. In these equations two potentials appear:
one potential is a generalized pressure. The second potential couples the 
magnetic and the flow shear components of the system. With the restriction to 
flux or at least line conserving flows one has to solve a modified Ohm's law. 
For the two dimensional case these are two coupled differential equations, which
represent the borderline between the resistive but flux conserving (or line 
conserving) case, and that of reconnective solutions. We discuss some 
simplified solutions of these equations.

\end{abstract}


\end{frontmatter}

\section{Introduction}

      A general problem of astrophysics, also reappearing in heliospheric physics, is the structure
      of the contact discontinuity which is located between the reverse shock\footnote{In
      heliospheric physics this is called termination shock.}
      of a stellar wind and the counterstreaming
      interstellar medium. The contact discontinuity is called the heliopause or more
      generally the astropause. Here not only two different flows are encountering, but also
      the electromagnetic fields transported by the flows.
      The aim of our work is to find magnetic field configurations in the vicinity
      of the stagnation point in the framework of MHD. 
      Similar work has already been done by \citet{Priest}. But their
      analysis used linearizations of the MHD--equations and their model
      was restricted to constant resistivities.

\section{Governing equations for stationary resistive MHD--flows in 2D}
   
Introducing the streaming vector $w=\sqrt{\rho}\,\vec v$
with $\vec\nabla\cdot\vec v=0$ and the electric potential $\phi_{e}(x,y)$
one can write the resistive MHD equations in the following form

 \begin{eqnarray}
\!\!\!\!\! \frac{\partial P}{\partial\zeta} &=& \Delta\zeta\qquad\,\,\,\frac{
 \partial S}{\partial\zeta}=w_{z}\qquad -\frac{\partial P}{\partial\alpha} = \frac{1}{\mu_{0}}\Delta
 \alpha \qquad -\frac{\partial S}{\partial\alpha}=\frac{1}{\mu_{0}}\, B_{z}\\[0.2em]
   \frac{\partial\left(\zeta,\alpha\right)}{\partial\left(x,y\right)} &=& 
   \frac{\partial\zeta}{\partial x}\,\frac{\partial\alpha}{\partial y} -
   \frac{\partial\zeta}{\partial y}\,\frac{\partial\alpha}{\partial x} =
   \sqrt{\rho}\left(R^{z} - E_{0}\right)\\[0.2em]
    -B_{z} w_{p} &=&  \sqrt{\rho}\left(R^{\zeta} + w_{p}\frac{\partial\phi_{e}}{\partial\zeta}\right)
     \quad\textrm{and}\quad 
     w_{z} B_{p} = \sqrt{\rho}\left(R^{\alpha} + B_{p}\frac{\partial\phi_{e}}{\partial\alpha}
\right)
     \end{eqnarray}

  where $\vec w = \vec\nabla\zeta\times\vec e_{z} + w_{z}\vec 
     e_{z}\equiv\vec w_{p} +  w_{z}\vec e_{z}$, $\vec B = \vec\nabla\alpha
     \times\vec e_{z} + B_{z}\vec e_{z}\equiv\vec B_{p} +  B_{z}\vec e_{z}$,
     $P\equiv p + \frac{1}{2\mu_{0}}B_{z}^2 + \frac{\rho}{2}\left(v_{x}^2 + 
v_{y}^2\right)$, and $S$ is a function of $x$ and $y$; $R^{\alpha}$, $R^{\zeta}$
and $R^{z}$ are the components of the generalized resistivity with Ohm's law
given by $\vec E + \vec v\times\vec B = \vec R$, and
$p$ is the plama pressure. The potentials $\zeta$ and $\alpha$ are functions of 
$x$ and $y$ only and are the streaming function respectively the magnetic 
flux function. The potential $S$ ensures the vanishing of the $z$-component of 
the Lorentz force, which guarantees the vanishing of the $z$-component of the 
pressure gradient and therefore that the equilibrium is two-dimensional.
Thus, the potential $S$ induces the appropriate coupling between $w_{z}$
and $B_{z}$. The equations for incompressible stationary resistive MHD in 2D 
had been derived by \cite{Thomas}, but without the relations for the components 
of the magnetic field and streaming vector in the invariant direction, which 
are derived in \citet{Nickeler}. Our procedure
follows mainly \citet{Thomas}, however, we admit here of the
$z$-component of the magnetic field and streaming vector.

To ensure the condition for magnetic field line freezing into the bulk plasma 
flow, one has to obey Ohm's law and the induction equation
\begin{equation}
\frac{\partial\vec B}{\partial t} - \vec\nabla\times\left( \vec v\times \vec B
\right) = \lambda\vec B\,.
\end{equation}
Derivations of this relation can be found in \citet{PriestForbes} and references
therein and in \citet{Hesse}.
For $\lambda = 0$ this reduces to the well-known frozen-in field condition
(flux conserving flow).
                                                                                
In the stationary two-dimensional case the above criterion must be written as
     \begin{eqnarray}
     &&\frac{1}{\sqrt{\rho}}\,\frac{\partial\left(\zeta,\alpha \right)}
     {\partial\left(x,y\right)} = \Lambda(\alpha)\,\,\textrm{and}\,\,
     \left(\frac{ \partial B_{z}}{\partial\alpha}
     + \frac{\partial w_{z}}{\partial\zeta} -\frac{w_{z}}{2\rho}\,
     \frac{d\rho}{d\zeta}  \right)\,\Lambda(\alpha) = \Lambda'(\alpha) B_{z}\label{zkompel}\\
     \Leftrightarrow &&\,\, \vec\nabla\times\left(\vec v\times\vec B \right)=
     \Lambda'(\alpha)\vec B\label{lico}
     \end{eqnarray}
     for the stationary two dimensional case. Equation (\ref{lico}) is equivalent to the critereon for
     line--freezing, which is given, e.g. by \citet{Vasyl}.
     \begin{figure}
     \begin{center}
          \includegraphics[width=8cm]{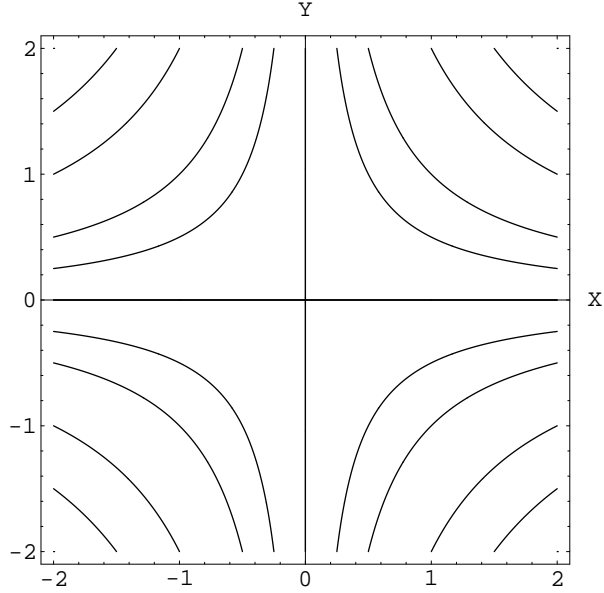}
       \end{center}
        \caption{Streamlines of a standard stagnation point of a symmetric
           astropause. The astropause is the y--axis.}
          \label{xpoint}
           \end{figure}
 \begin{figure} 
  \begin{center}
            \includegraphics[width=8cm]{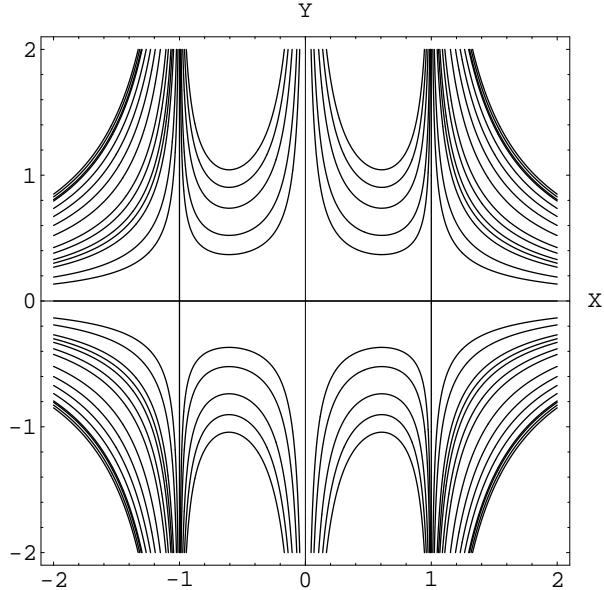}
       \end{center}
         \caption{Magnetic field lines resulting from solving Ohm's law with
          respect to the velocity field of Fig.\,1 for $E_{0}\neq 0$; for $E_{0}=0$ the magnetic field
          would be exactly flow--aligned, if the shear components are zero.}
       \end{figure}
        The term $\Lambda(\alpha)$ is the difference between the resistive term
        and the constant electric field $E_{0}\neq 0$ in $z$--direction\footnote{$E_{0}=0$
        would imply that the flow is purely field aligned}.
        A simple case is $\Lambda(\alpha)= constant=\Lambda_{0}=-E_{0}$ which neglects
        resistive terms. Assuming a standard stagnation point given by the flow field
        $\zeta=axy$, which is shown in Fig.\,1, we are able to calculate a
        wide class of solutions of ideal Ohm's law
        for the magnetic field by using the theory of characteristics.
        These solutions of ideal Ohm's law are given by 
         \begin{equation}
        \alpha= -\frac{E_{0}}{2a}\,\sqrt{\rho}\left(k_1\,\ln{\left(x/x_{0}\right)^2} +
        \left(1+k_1\right)\,\ln{\left(y/y_{0}\right)^2}\right)
        \end{equation}
        Field lines of a special solution of the Ohm's law around a
        standard stagnation point can be seen in Fig.\,2.
        where in this special case the constants $a$, $x_{0}$ and $y_{0}$ are set to 1 and $k_{1}$
        is set to -1, which could represent the magnetic fields in the vicinity 
        around a stagnation point of an astrosphere, where the flow vector $\vec w$ 
        vanishes at the ``nose'' (i.e. stagnation point) of the astropause.
        However, in the case of ideality ($\Lambda=const.$) the velocity field diverges,
        although $|\vec w|$ remains finite, so that the ``stagnation point'', is only a
        ``topological'' stagnation point. Only for the field $\vec w$ the origin is
        a stagnation point.
        Here we have to choose the density to be zero on the separatrices to avoid magnetic
        singularities. 
        Therefore around the separatrices there must be a kind of
        a ``vacuum gap'' to ensure that the mass current density vector $\sqrt{\rho}\vec w$
        remains finite. 
         This leads to a double separatrix 
        structure of the magnetic field (see Fig.\,2). So the constraint of ``idealness''
        leads to a magnetic structure which is more complex then the underlying structure 
        of the flow field.

        \section{Problems and Discussion}

         The class of solutions we found does not ensure that these are also solutions
         of the Euler equation. It seems to be more probable, that stationary MHD equilibria,
         conserving magnetic flux, do not exist close to the
         standard stagnation point, at least not without other forces, e.g. pressure anisotropies.
         It seems that to date only for a constant resistivity 
         two dimensional annihilation solutions (without shear components) have been found
          \citep{Thomas}.
         So our ansatz gives a possibility for further investigations, to find non
         flux conserving, but line conserving solutions $\Lambda=\Lambda(\alpha)$, so called
         annihilation solutions, as magnetic flux is annihilated.




\end{document}